\documentclass[11pt]{article}
\usepackage{moriond,epsfig,amssymb}

\def\Journal#1#2#3#4{{#1} {\bf #2}, #3 (#4)}

\def\be{\begin{equation}}
\def\ee{\end{equation}}
\def\bea{\begin{eqnarray}}
\def\eea{\end{eqnarray}}

\begin{document}
\title{Statistics of quantum transport in metal nanowires with surface disorder}

\author{\underline{J\'er\^ome B\"urki}$^1$,
Charles A. Stafford$^2$}

\address{$^1$Lyman Lab of Physics, Harvard University, Cambridge MA 02138, USA}
\address{$^2$Department of Physics, University of Arizona, 1118 E.\ 4th St.,
Tucson AZ 85721, USA}

\maketitle\abstracts{
Experimental conductance histograms 
built from several thousand successive breakings of sodium 
nanowires \cite{yanson99,yanson00} exhibit peaks up to rather high conductance 
values ($\sim 100\times 2e^2/h$).
In this paper, we present results from a disordered free-electron model of a 
metallic nanowire, which was previously successful
in describing both conductance histograms and shot noise measurements in
gold nanocontacts with much lower
conductances.\cite{burki99} We find in particular that, with a modification of 
the model of disorder, the conductance 
histogram can be understood as an interplay of conductance quantization and 
disorder for low conductances 
($G \lesssim 10\times 2e^2/h$), 
while peaks corresponding to higher conductance are actually a combination of 
several ``quantized conductance peaks''.
We also predict a saturation of the shot noise at high conductance to about 
$1/10$ of its classical value $2e\bar{I}$.
}

\section{Introduction}

Recent experiments by Yanson {\sl et al.} \cite{yanson99,yanson00} measuring 
conductance histograms
for sodium nanocontacts have revealed the persistance of well-defined 
conductance peaks up to rather high 
conductance values, 
of order $100G_0$, where $G_0=2e^2/h$ is the quantum of conductance. 
A semi-classical calculation of Kassubek {\sl et al.} \cite{kassubek00}
provided a description of the histograms in terms of the stability of 
nanowires of certain radii, but the shift of the 
experimental peaks with respect to quantized conductance values makes it 
difficult to identify a peak with a particular quantized 
conductance. 
We therefore solve in this paper a free-electron model \cite{stafford97a}
of a metallic nanowire that includes disorder
\cite{burki99} and allows for a more quantitative identification of 
the peaks, as well as the calculation of other quantities such as shot noise.

The sequence of conductance peaks observed \cite{yanson99,yanson00}
 at low temperature 
($1,3,5,6,\dots\times G_0$) is suggestive of 
cylindrical symmetry. We therefore use a three-dimensional model of a 
cylindrical wire with a constriction, presented
in section \ref{sec:model}. The extension of the recursive Green's function 
numerical method \cite{burki99} to three-dimensional 
contacts is discussed in section \ref{sec:method}. 
Finally, the results are presented and discussed in section \ref{sec:results}.

\section{Model\label{sec:model}}

As in previous papers,\cite{burki99,stafford97a} we use a free-electron model 
of a metallic nanowire. The nanowire is taken to be
an infinite cylinder of radius $R$ with a finite deformable part representing 
the contact, defined by the function 
$r(z)=R_{min}+(R-R_{min})\left[3u^2-3u^4+u^6\right]_{u=2z/L}$, where 
$R_{min}$ is determined at each step of the elongation $L$
such that the volume of the contact is kept constant. Independent electrons 
are confined within
the wire by hard-wall boundary conditions and the Fermi energy of the electron 
gas is taken to be the bulk sodium Fermi energy,
namely $\varepsilon_F = 3.23 eV$. All calculation are made at zero temperature.

The bulk model of disorder used previously to model gold nanocontacts
\cite{burki99} is problematic for sodium
because of the larger size \cite{yanson99,yanson00} of the contacts:
A disorder resistance of a few hundred Ohms is necessary to 
shift the lower conductance peaks back onto quantized values, corresponding 
to a conductance of less than $100G_0$. 
If the disorder resistance is to be constant during elongation of the contact, 
which is the case with bulk disorder, it would be impossible to measure a 
conductance larger than this value, which is clearly
not the case.\cite{yanson99}
We therefore need a model of disorder whose resistance decreases with 
increasing conductance of the contact.

The material used in experiments being of very high purity, 
the main source of backscattering is likely to be surface roughness
\cite{bratkovsky} or surface defects. 
We therefore introduce the following model of surface disorder: 
(i) As previously,\cite{burki99} we use $\delta$-function impurities of fixed 
strength $W$ and random positions; 
(ii) The initial positions of the impurities are restricted to a length 
$L_{dis}$ of the conductor that includes the 
constriction, and within a distance $d$ of the surface of the wire 
(i.e. $R-d \leqslant r \leqslant R$, $r$ being the radial coordinate of the impurity);
(iii) During elongation of the contact, the distance of the impurity to the 
surface is kept constant, while its longitudinal position
is changed such that the linear density of impurities is constant along the 
wire.

With this model, the resistance due to impurities will increase as the minimal 
cross-section area decreases. The thickness $d$ of the
disordered shell is chosen on physical grounds to be essentially one atomic 
layer, or $k_F d = 3$. 
We want to emphasize that this surface disorder model is likely to be a better 
model for gold as well, but all conductance 
histograms published for gold \cite{costa-kramer97c} were restricted to 
conductances lower than $10G_0$, 
in which regime both models give similar results.
 
\section{Method\label{sec:method}}

In principle, the recursive Green's function method used in our earlier paper 
\cite{burki99} can be extended without further 
complication to a three-dimensional model such as considered in this work. 
In practice, however, one would need a very 
fine discretization of space in order to reproduce the correct degeneracies of 
the transverse energy levels, 
making the computation intractable, especially for the larger contacts we want 
to consider.
Instead, we discretize only the longitudinal coordinate $z$ after the change 
of coordinates that brings the
constriction back to a cylinder, 
and use a basis of transverse eigenstates of the ``free'' problem 
(i.e. without a constriction or disorder) 
to get a matrix representation of the Hamiltonian of a slice. 
In order to have a finite matrix, we introduce an energy cutoff, 
which is chosen such that the number of states considered is of the order of 
twice the number of open channels.
Except for this change of basis and the use of a three-dimensional model, the 
method is a straightforward generalization 
of our previous work.\cite{burki99} 
Once the transmission matrix $t(\varepsilon_F)$ at the Fermi energy has been 
obtained from the Green's
function,\cite{burki99} the conductance $G$ and shot noise $S_I$ at zero 
temperature are computed using the Landauer-type formulae

\begin{equation}
G = \frac{2e^2}{h}\mbox{Tr}\left(t^{\dagger}t\right),\qquad\qquad
S_I = 2e\bar{I} \frac{\mbox{Tr}\left[t^{\dagger}t\left(1-t^{\dagger}t\right)
\right]}{\mbox{Tr}\left(t^{\dagger}t\right)}.
\label{eq:GandSi}
\end{equation}

In order to generate a histogram of conductance, we compute the conductance as 
a function of elongation for a given configuration
of disorder, divide the conductance axis in intervals of size
 $\Delta G=0.1G_0$ and count the number of points in each 
interval. We repeat this for 300 configurations of disorder, 
averaging at the same time on the geometry of the contact by using
different initial lengths $L_0$ for the constriction 
($5 \leqslant k_F L_0 \leqslant 25$). We then smoothen the histogram by taking an average
over three bins with relative weights $1,2,1$, 
similarly to what is done in the experiments.\cite{yanson99}
We can also resolve individual ``quantized'' 
conductance peaks by restricting the elongation 
to intervals corresponding to a fixed number of open channels of
the clean adiabatic wire.
As to the shot noise, we average its value at a given conductance, 
using the same set of data as for the histogram.

\section{Results\label{sec:results}}
\subsection{Conductance histogram}

\begin{figure}[h]
\psfig{figure=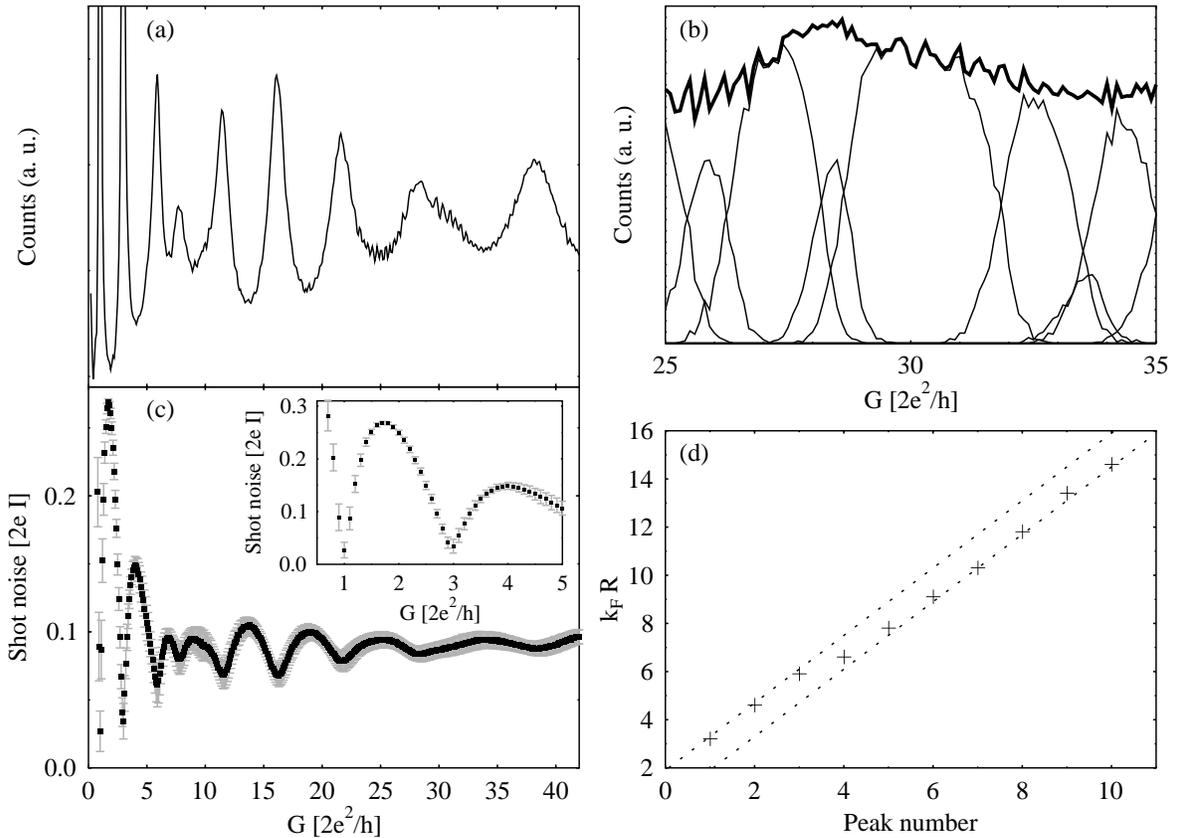,height=11.75cm}
\caption{(a) Conductance histogram built from 300 disorder configurations.
(b) Blowup of the eighth peak of the 
histogram (thick line). 
The thiner lines represent contributions to the histogram
from individual ``quantized'' conductance peaks.
(c) Average shot noise of the contact (squares) together with the 
standard deviation (bars) as a function of the 
conductance of the contact; the inset blows up the graph at small conductances.
(d) Relation between the radius of the wire and the peak number (crosses); 
the dotted lines are a guide to the eye.
\label{fig}}
\end{figure}

Following the procedure described above, we generated the histogram of
Fig.\ \ref{fig}(a).  The first two peaks, near $G_0$ and $3G_0$,
can be thought of as individual quantized conductance peaks, 
broadened and shifted downward slightly by disorder.  However, the 
third peak is a mixture of $5G_0$ and $6G_0$, and higher peaks are typically
even more complicated admixtures.
In Fig.\ \ref{fig}(b), the peak near $28G_0$ is shown, along with
its decomposition in terms of individual 
quantized components; one sees that it is actually composed of many individual
peaks.  The peaks in the conductance histogram are thus generically not
ascribable to individual quantized conductance channels, but rather
represent a gross shell structure.\cite{yanson99,yanson00}
One notes that the agreement with the experimental results of 
Yanson {\sl et al.}\cite{yanson99,yanson00} 
is reasonable for conductances
lower than $30G_0$, except for the one peak around 
$7G_0$, which seems to be out of place, and which will be
discussed later.
For larger conductance, the peaks start to be out of phase with the
experimental peaks, which might have several causes:
First of all, the disorder present in the experimental nanowires
might be stronger or weaker than that in our model.
%
Secondly, our calculation is made at zero temperature, 
while the experiment is made around $90 K$, where surface 
diffusion of atoms allows the contact to find more stable configurations, 
so that those peaks might reflect the stability of the contact, 
rather than quantum transport effects.

Coming back to the peak near $7G_0$, 
the discrepancy of its position and its relative weakness suggests that it 
might actually
disappear if a larger amount of disorder was used. In that case, the higher 
conductance peaks would be shifted further downward, and would presumably
correspond to lower experimental peaks.
This hypothesis is supported by Fig. \ref{fig}(d), which shows the radius of 
the wire, as obtained from the Sharvin formula
$G_s=\left(k_F R/2\right)^2\left(1-2/k_F R\right)$, 
as a function of the peak number. As is suggested by the two dotted lines, 
corresponding to a shift of the peak number by one, 
the linear relation between the radius and the peak number would be
better if the fourth peak, which is the peak in question, were removed.
The hypothesis of stronger disorder, currently under investigation, 
seems promising.

\subsection{Shot noise}

The shot noise, computed as described in Sec.\ 
\ref{sec:method}, is shown in Fig.\ \ref{fig}(c). It shows a strong suppression
at low quantized-conductance values and weaker minima corresponding to 
higher peaks in the conductance histogram, as 
could be expected from our previous results on gold nanocontacts.\cite{burki99}
At larger conductance values, we observe a saturation of the shot noise at 
about $1/10$ of the classical value $2e \bar{I}$, 
in contrast to a random-matrix prediction by Beenakker and 
Melsen \cite{beenakker94} of a suppression factor of $1/3$ 
in a disordered nanowire. 
Part of this discrepancy may come from the difference in disorder 
(surface versus bulk), but even
our results for gold, which used a model much more similar to the one of 
Ref.\ \cite{beenakker94}, showed a saturation value that was about
one half of the random-matrix prediction. 
The reason for this discrepancy is not fully understood, but it may 
indicate that tunneling through the nanocontact, which is neglected
in the model of Ref.\ \cite{beenakker94}, plays an essential role in the 
shot noise in the saturation region, just as it does in the ballistic
region at small conductance.


\section{Conclusions}

We have presented a model of surface disorder in a free-electron nanowire 
whose numerical solution, using a variation of
the Green's function technique presented in our previous paper,\cite{burki99} 
helps understand conductance histograms
obtained experimentally for sodium nanowires, and provides a prediction for 
what the shot noise of such wires should look like.
Our results show a discrepancy for the saturation value of the shot noise 
compared to random matrix theory \cite{beenakker94}
that is so far not understood.

\section*{Acknowledgments}
JB was supported by a fellowship of the Swiss National Science 
Foundation.  CAS was supported by NSF Grant DMR0072703.

\end{document}